\documentclass[prl,twocolumn,showpacs,superscriptaddress]{revtex4}
\usepackage{dcolumn}
\usepackage{bm}
\usepackage{graphicx}
\usepackage{amsmath}
\usepackage{amssymb}
\usepackage{bm}
\begin{document}
\title{A precursor state to unconventional superconductivity in
CeIrIn${_5}$}
\author{Sunil Nair}
\affiliation{Max Planck Institute for Chemical Physics of Solids,
Noethnitzer Str. 40, 01187 Dresden, Germany.}
\author{S. Wirth}
\affiliation{Max Planck Institute for Chemical Physics of Solids,
Noethnitzer Str. 40, 01187 Dresden, Germany.}
\author{M. Nicklas}
\affiliation{Max Planck Institute for Chemical Physics of Solids,
Noethnitzer Str. 40, 01187 Dresden, Germany.}
\author{J. L. Sarrao}
\affiliation{Los Alamos National Laboratory, Los Alamos, New
Mexico 87545, USA.}
\author{J. D. Thompson}
\affiliation{Los Alamos National Laboratory, Los Alamos, New
Mexico 87545, USA.}
\author{Z. Fisk}
\affiliation{University of California, Irvine, California 92697, USA.}
\author{F. Steglich}
\affiliation{Max Planck Institute for Chemical Physics of Solids,
Noethnitzer Str. 40, 01187 Dresden, Germany.}
\date{\today}
\begin{abstract}
We present sensitive measurements of the Hall effect and
magnetoresistance in CeIrIn${_5}$ down to temperatures of 50 mK
and magnetic fields up to 15 T. The presence of a low temperature
coherent Kondo state is established. Deviations from Kohler's
rule and a quadratic temperature dependence of the cotangent of
the Hall angle are reminiscent of properties observed in the high
temperature superconducting cuprates. The most striking
observation pertains to the presence of a \textit{precursor}
state---characterized by a change in the Hall mobility---that
appears to precede the superconductivity in this material, in
similarity to the pseudogap in the cuprate high $T_c$
superconductors.
\end{abstract}
\maketitle

The phenomenon of heavy fermion (HF) superconductivity (SC)
continues to be a central focus of investigations into strongly
correlated electron systems. The initial interest in these
systems was primarily centered on reconciling the observation of
SC in an inherently magnetic environment and its interplay with
the effect of Kondo screening in a correlated Fermi liquid.
However, its re-emergence has been dramatic, with current
emphasis being placed on understanding the phenomenon of magnetic
quantum critical points (QCP). Here, a QCP refers to a zero
temperature ($T$) magnetic instability which can be tuned by a
non-thermal control parameter like magnetic field ($H$), pressure
($P$) or composition. Such a QCP is believed to crucially
influence physical properties in a large region of the
\textit{H}--\textit{T}--\textit{P} phase space in its vicinity.
The added incentive is to bridge our understanding of the HF
systems and the high-temperature superconductors; two distinctly
separate classes of systems where SC and magnetism are intricately
connected.

The more recent discovery of the Ce$M$In${_5}$ (where $M$: Co, Rh
or Ir) family of HF systems has further enriched this field
\cite{petrovic1}. These layered materials crystallize in the
tetragonal structure, and the quasi two-dimensional character of
their Fermi surfaces (FS) was confirmed by de Haas-van Alphen
(dHvA) measurements \cite{settai}. Moreover, both the
superconducting and normal states in these materials have been
reported to be highly unusual. For instance, in CeCoIn${_5}$
(with the highest reported ambient pressure superconducting
transition temperature $T{_c}$ $\approx$ 2.3 K among Ce-based HF
systems \cite{petrovic1}) measurements of specific heat, thermal
conductivity and Andreev reflection have indicated\cite{matsuda}
that the superconducting gap function has line nodes, and is most
likely to have a \textit{d}-wave symmetry. Coupled with other
observations like a linear \textit{T} dependent resistivity and a
strongly \textit{T} dependent Hall coefficient $R{_H}$, a
remarkable similarity of these systems with the high $T{_c}$
cuprates was suggested \cite{nakajima}.

CeIrIn${_5}$ is the other ambient pressure superconductor in this
series, with a bulk \textit{T}${_c}$ $\approx$ 0.4 K and a
resistive \textit{T}${_c}$ $\approx$ 1.2 K \cite{petrovic2}. In
spite of a band structure similar to its Co and Rh counterparts,
striking differences have been observed in both its
superconducting and normal state properties. The primary
difference pertains to the position of the magnetic instability
with respect to the SC region in these systems. In CeCoIn${_5}$,
the magnetic field tuned QCP appears to be close to the upper
critical field $H_{c2}(T)$ of the superconductor
\cite{paglione,bianchi,singh}. However, in CeIrIn${_5}$ the
magnetic instability is reported to lie far away from the
superconducting region as has been inferred from prior
investigations using \textit{H}, \textit{P} and chemical
composition as control parameters; with \textit{H} suppressing
rather than enhancing the Landau Fermi liquid (FL) state
\cite{capan,nicklas}. This has also led to suggestions that
CeIrIn${_5}$ may be a prospective system, in which SC is mediated
by charge valence fluctuations \cite{holmes}. Unlike its Co
counterpart, the $H$--$T$ phase space in the vicinity of the SC
region in CeIrIn${_5}$ is expected to be free from the influence
of the magnetic instability, thus enabling a cleaner
investigation of the superconducting state in this HF system. In
this letter, we report the investigation of CeIrIn${_5}$ using
sensitive magnetoresistance and Hall effect measurements. Besides
the observation of a low-temperature coherent Kondo state,
experimental signatures of the presence of a \textit{precursor}
state that envelops the superconducting region in this system is
seen---an observation which could imply that the condensation of
electrons into Cooper pairs is preceded by an electronic state
hitherto unexplored in this class of materials.

The magnetotransport measurements were mainly conducted as
isothermal field sweeps on high quality single crystals of
CeIrIn${_5}$ (resistivity $\rho$ $\approx$ 1.75 $\mu\Omega$ at
1.35 K), with the crystallographic \textit{c} axis parallel to
\textit{H} and the current of $\approx$ 20 $\mu$A being applied
along the \textit{ab} plane. In addition, temperature sweeps were
carried out at selected $H$ to complement the typically more
sensitive isothermal measurements. The set up is based on
\cite{singh}, with additional low-noise preamplifiers used to
enable a sensitivity of the order of $\pm$0.01 nV.  The Hall
voltage is obtained as the asymmetric component under field
reversal.

\begin{figure}[tb]
\centering \includegraphics[width=7.4cm,clip]{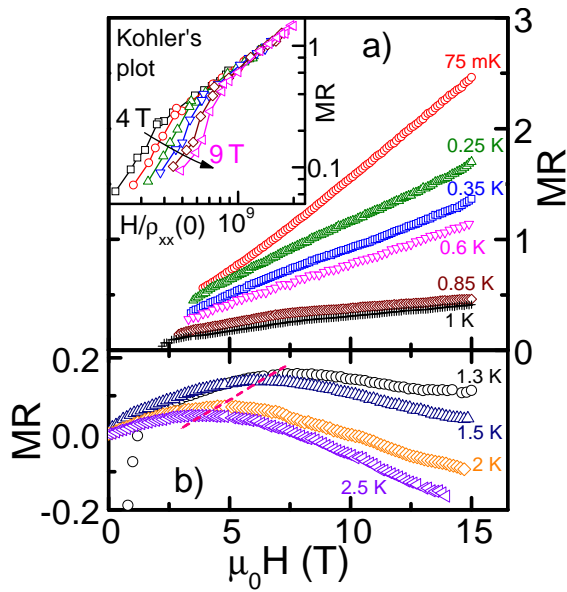}
\caption{Magnetoresistance, MR
$=\Delta\rho{_{xx}}$/$\rho{_{xx}}(0)$, as measured in
CeIrIn${_5}$ at constant $T \le$ 1 K (a) and $T >$ 1 K (b). The
maximum in MR indicates the coherent transition, dashed line in
(b). Inset to (a): Violation of Kohler's scaling rule in the NFL
regime. In this plot, $T$ is an implicit parameter, and the
abscissa is given in units of T/$\Omega$m.} \label{fig1}
\end{figure}

The magnetoresistance MR = [$\rho{_{xx}}(H)-\rho{_{xx}}(0)]/
\rho{_{xx}}$(0) = $\Delta\rho{_{xx}}$/$\rho{_{xx}}(0)$ as
measured in CeIrIn${_5}$ at selected temperatures is shown in
Fig.~\ref{fig1}. At the lowest measured $T$, the MR is seen to be
positive and subquadratic as a function of $H$. With increasing
\textit{T}, a negative contribution to the MR is seen to arise at
high \textit{H}, with competition between the negative and
positive contributions eventually resulting in a
\textit{crossover}, where the sign of $\partial($MR$)$/$\partial
H |_T$ changes. This crossover can be identified to be the
\textit{coherent} transition \cite{brandt}, marking the onset of
the dense Kondo regime at $H{_{coh}}$ [dashed line in
Fig.~\ref{fig1}(b)]. Since the negative component of MR stems
from the suppression of spin flip scattering, it is expected to
grow with increasing $T$. This should result in the crossover
moving to lower \textit{H} with increasing \textit{T}, as is
indeed observed in our data. These measurements are in agreement
with prior optical conductivity measurements which indicate the
formation of a low-$T$ coherent state in CeIrIn${_{5}}$
\cite{optical}. A recent two fluid description of the Kondo
lattice has suggested that the \textit{T}$\rightarrow$0 ground
state in these materials can be described by a sum of the (single
Ce$^{3+}$ ion) Kondo gas and a coherent Kondo liquid, with the
latter being about 95$\%$ of the whole in the case of
CeIrIn${_5}$ \cite{fluid}. In the high-$T$ limit, this Kondo
coherence would be expected to form below $T \approx 20$ K
\cite{fluid}. In the low-$T$ limit, this coherence scale of the
Kondo lattice is anticipated to vanish at the magnetic
instability ($\gtrsim 25$ T, \cite{take,kim}).

In the FL description, the low-$T$ positive MR arises from the
bending of the electron trajectory by the Lorentz force. Assuming
isotropic scattering times at all points on the FS, the
\begin{figure}[tb]
\centering \includegraphics[width=7.4cm,clip]{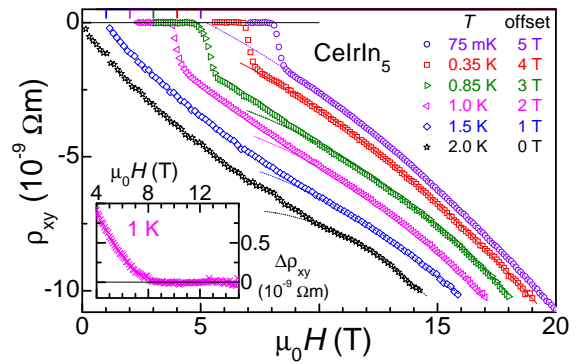}
\caption{Isothermal Hall resistivity $\rho{_{xy}}(H)$. For
clarity, each data set is offset along the $H$ axis, with the
offset value indicated in the legend. The dotted lines (guides to
the eye) indicate the quadratic \textit{H} regime which shifts to
higher fields with increasing $T$. Inset: The difference
$\Delta(\rho{_{xy}})$ between measured values $\rho{_{xy}}$ and
the quadratic fit, for $T = 1$ K.} \label{fig2}
\end{figure}
MR is expected to scale as a function of $H/\rho{_{xx}}$(0). This
is known as Kohler's rule \cite{pippard}, and should hold
regardless of the topology and the symmetry of the FS. The inset
of Fig.~\ref{fig1}(a) exhibits a Kohler's plot for CeIrIn${_5}$,
clearly indicating a violation of Kohler's rule. The deviation
from scaling occurs in the non Fermi liquid (NFL) regime, and the
$T$ and $H$ dependences of this transition match well with prior
reports \cite{capan}.

The Hall effect, a rather complex quantity, has proven to be of
great significance in the investigation of HF systems in the
vicinity of a QCP \cite{paschen}. This is due to the fact that in
HF systems, the low temperature $R{_H}$ predominantly arises from
the normal part of the Hall effect \cite{fert}, and thus can be
used to monitor the evolution of the FS volume. The results for
the Hall resistivities $\rho_{xy}(H)$ in CeIrIn${_5}$ at selected
$T$ are shown in Fig.~\ref{fig2}. The measured $\rho{_{xy}}$ is
seen to be negative (indicating electron-dominated transport) and
nonlinear in \textit{H} down to the lowest measured \textit{T}.
Their magnitudes are in good agreement with prior
high-temperature data \cite{hundley}. At low \textit{T},
$\rho{_{xy}}$ exhibits a nearly quadratic \textit{H} dependence,
with this quadratic regime only valid for higher fields as
\textit{T} is increased. It is interesting to note that in spite
of the complex band structure of CeIrIn${_5}$, the observed
$\rho{_{xy}}$ behavior can---at least qualitatively---be
explained on the basis of that expected for simple compensated
metals. Here, a quadratic $H$ dependence is anticipated
\cite{hurd} in the high field limit, i.e., when $\omega{_c}\tau
\gg$ 1 ($\omega{_c} = eH /m^*$ is the cyclotron frequency,
$m{^*}$ is the effective mass and $\tau$ the average time between
scattering events). Since CeIrIn${_5}$ {\it is} a compensated
metal, as concluded from dHvA measurements \cite{haga}, this
observed $H$ dependence is not unexpected. A decrease in $\tau$
with increasing \textit{T} explains the shift of the quadratic
regime to higher fields at higher \textit{T}. In this context we
\begin{figure}[tb]
\centering \includegraphics[width=8.2cm,clip]{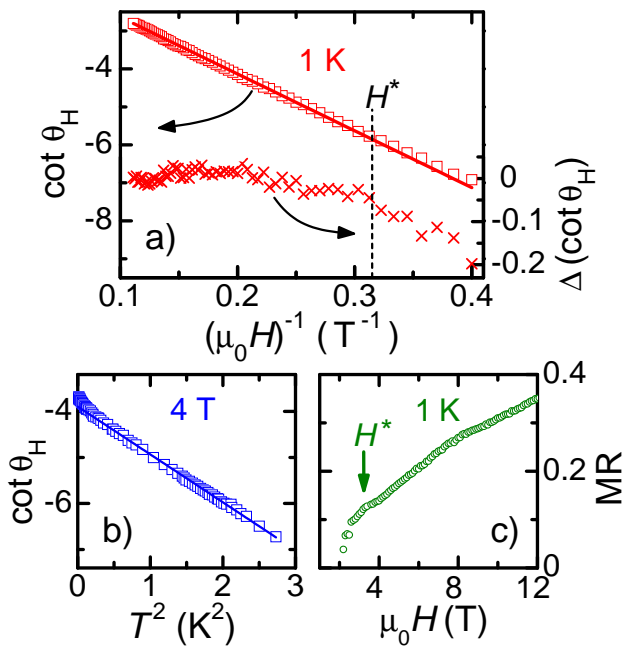}
\caption{(a) $H^{-1}$ dependence of $\cot\theta{_H}$ at $T = 1$ K.
Experimental data $\cot\theta{_H}$ deviate from a linear $H^{-1}$
dependence below $H{^*}$. The difference $\Delta(\cot\theta{_H})$
($\times$ and right scale) between $\cot\theta{_H}$ and a linear
fit to $\cot\theta{_H}$ is used to identify $H{^*}$. (b) $T{^2}$
dependence of $\cot\theta{_H}$. (c) \textit{H} dependence of the
MR at 1 K, with the anomaly at $H{^*}$ marked by an arrow.}
\label{fig3}
\end{figure}
note that in spite of the reported similarity of the Ir and Co
based systems with respect to their band structure, there are
obvious differences in $\rho{_{xy}}$ as measured in CeIrIn${_5}$
with that of its Co counterpart reported earlier \cite{singh}.
For instance, in CeCoIn${_5}$ $\rho{_{xy}}$ was linear at the
lowest \textit{T} and a (\textit{P} dependent) signature in
\textit{R}${_H}$ was observed which was attributed to arise as a
consequence of critical spin fluctuations. A likely reason for
this behavior \textit{not} being observed in CeIrIn${_5}$ could
be that the $H$--$T$ phase space explored by our measurements
does not encompass the putative QCP, a consequence of the fact
that the magnetic instability in each system lies in very
different regions of the $H$--$T$ phase space.

The cotangent of the Hall angle ($\cot\theta{_H} = \rho{_{xx}}/
\rho{_{xy}}$) is directly related to the charge carrier mobility,
and is a quantity of fundamental interest \cite{ong}. In many
systems including the cuprates, it has been observed to vary as
\textit{T}${^2}$. Since $\rho{_{xx}}$ in cuprates is observed to
be linear in \textit{T}, this functional form of
$\cot\theta{_{H}}$ reflects a Hall scattering rate
($\tau{_{H}}^{-1}$) which is at variance with the scattering rate
($\tau{_{tr}}^{-1}$) governing the resistivity.
Fig.~\ref{fig3}(b) exhibits the \textit{T}${^2}$ dependence of
$\cot\theta{_H}$ as deduced in CeIrIn${_5}$, a behavior observed
in a substantial region of the $H$--$T$ phase space.
Interestingly, however, systematic deviations are seen at low
\textit{T}. Though this aspect has not been addressed in the
context of HF systems, such deviations from $T^2$ have been used
to mark the onset of the pseudogap phase in some high
\textit{T}${_c}$ cuprates \cite{abe}. Our measurement protocol
enables us to evaluate in more detail the \textit{H} dependence
of this quantity, and careful inspection of the
\textit{H}-\textit{T} phase space in the vicinity of the SC
region shows that $\cot\theta{_H}$ has a \textit{H}${^{-1}}$
dependence. Interestingly, as one decreases \textit{T} and
approaches the superconducting region, systematic deviations from
this \textit{H}$^{-1}$ dependence are observed below a critical
field $H^*$. This is shown for the example of $T = 1$ K in
Fig.~\ref{fig3}(a). The difference $\Delta(\cot \theta_H)$ of the
experimental data $\cot \theta_H$ from a linear fit ($\times$ and
right axis) is used to identify $H^*$ below which
$\cot\theta{_H}$ deviates from $H{^{-1}}$. We emphasize that this
deviation is also reflected as subtle feature in the $H$
dependence of the MR, see Fig.~\ref{fig3}(c).

Attempts to reconcile the observed functional form of
$\cot\theta{_{H}}$ in cuprates with theory have primarily been
based on (i) a model within the Luttinger liquid formalism, which
relates the different scattering rates to distinct particles with
dissimilar scattering events \cite{anderson} and (ii) a nearly
antiferromagnetic (AF) FL description, which predicts anisotropic
scattering on the FS, with $\tau{_{H}}^{-1}$ and
$\tau{_{tr}}^{-1}$ being dictated by scattering events on
\begin{figure}[tb]
\centering \includegraphics[width=7.0cm,clip]{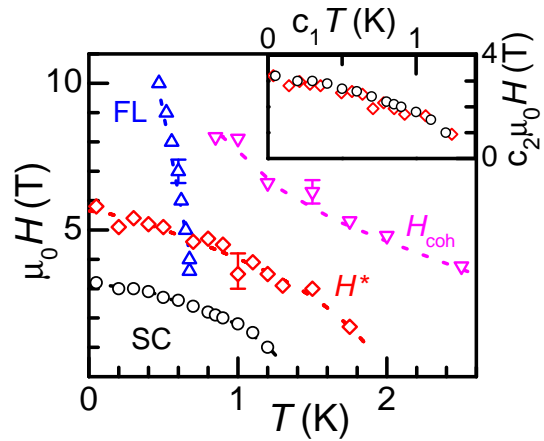}
\caption{\textit{H}-\textit{T} phase diagram of CeIrIn${_5}$
determined from a combination of Hall effect and MR measurements
(lines are guides to the eye). Inset: Scaling of $H^*(T)$ with
$H_{c2}(T)$, with $c_1 = c_2 =$ 1 for $H_{c2}(T)$ and $c_1 =$ 0.7,
$c_2 =$ 0.55 for $H^*(T)$.} \label{fig4}
\end{figure}
different parts of the FS \cite{pines}. In Anderson's theory
\cite{anderson}, the Hall angle is governed only by
$\tau{_{H}}^{-1}$ and can be expressed as $\cot\theta{_{H}} = 1/
\omega{_c}\tau{_H}$. Thus, $\cot\theta{_{H}}$ would be expected
to vary as $H^{-1}$, with the slope being a function of
$\tau{_H}$${^{-1}}$; as is observed in our case. Fig.~\ref{fig4}
shows the $H$--$T$ phase diagram of CeIrIn${_5}$, with the FL--NFL
transition as determined from deviations from Kohler's rule
(inset of Fig.~\ref{fig1}), the onset of Kondo coherence
determined from the maximum in MR, Fig.~\ref{fig1}(b), and the
deviations from \textit{H}${^{-1}}$ at $H{^*}$
[Fig.~\ref{fig3}(a)] clearly marked out. The most striking
feature here is the envelope of $H^{*}$ around the
superconducting region, indicating that the condensation of
itinerant electrons into Cooper pairs in CeIrIn${_5}$ is preceded
by a \textit{precursor} state associated with a change in the
Hall mobility. Moreover, the critical field $H^{*}(T)$ of this
precursor state can be scaled onto $H_{c2}(T)$ (as shown in the
inset) suggesting that both these states might arise from the
same underlying mechanism.

For the system CeCoIn${_5}$, a precursor state has also been
deduced from thermopower and Nernst effect measurements
\cite{onuki}. In analogy with the cuprates, a vortex-liquid state
was suggested, where thermal phase and vortex fluctuations result
in short-range phase coherence \cite{ong2}. Though this cannot be
ruled out as the cause of our experimental observations, we note
that we have failed to observe a measurable Hall signal in the
mixed state of CeIrIn${_5}$, probably indicating that vortex
dynamics is rather weak. Moreover, prior investigations have
failed to reveal a diamagnetic response in this phase space
region.

An alternative scenario would involve a strong anisotropy of the
transport scattering rates, which in turn arise from the coupling
of AF fluctuations to the (otherwise isotropic) FL formalism.
This is achieved by the formation of hot (and cold) spots on
different regions of the FS. Here, hot spots refer to positions
on the FS surface, where the AF Brillouin zone intersects it, and
the electron lifetimes are very short. Thus, all the transport
coefficients would be renormalized with respect to the ratio
$\tau{_{cold}}/\tau{_{hot}}$, reflecting the anisotropy of the
FS. An increasing \textit{H} would be expected to suppress these
AF fluctuations, thus effectively \textit{closing} the gapped
regions of the FS. It is to be noted that transport \cite{sid}
and (\textit{P} dependent) nuclear quadrupole resonance (NQR)
measurements \cite{kawasaki} have been used to speculate on the
presence of a pseudogap phase in the Co and Rh counterparts,
respectively. A related scenario was recently reported: an
\textit{anisotropic} destruction of the FS in CeCoIn${_5}$ in the
\textit{T}$\rightarrow$0 limit, reminiscent of the pseudogap
phase in the cuprates \cite{tanatar}. In spite of the absence of
a magnetic instability in the immediate vicinity of the SC region
in CeIrIn${_5}$ (in contrast to its Co counterpart), a prior NQR
study has inferred on the presence of anisotropic spin
fluctuations in CeIrIn${_5}$ \cite{zheng} indicating that a
similar mechanism could be at play in this system.

In the absence of other experimental evidences, one can only
speculate on the nature of low-lying electronic excitations which
give rise to this precursor state. It may arise as a consequence
of AF fluctuations as discussed above, or may even signify a
hitherto unknown form of unconventional order. The scaling of
$H^{*}(T)$ with $H{_{c2}}(T)$ is striking, and points towards a
common origin of the precursor state and the SC in this system.
The FL--NFL crossover in the phase diagram is related to the
presence of the magnetic instability at $\mu_0 H\approx 25$ T.
This instability would also be expected to influence
$H_{\text{coh}}$, and a crossing between the FL and
$H_{\text{coh}}$ lines is improbable. However, it is pertinent to
note that the precursor state encompasses both the FL and NFL
regimes, and is suppressed to $T$$\rightarrow$0 by the applied $H$
in the FL regime of the phase diagram. This is in contrast to what
is observed in the cuprates implying that theoretical approaches
commonly employed in the latter may have only limited
applicability in this case. The low-$T$ phase diagram of
CeIrIn${_5}$ is clearly dictated by both, the magnetic
instability as well as the presence of the precursor state.
Whether they \textit{complement} or \textit{compete} with each
other, is an aspect which more direct experiments would need to
resolve.

In summary, Hall effect and MR measurements clearly demarcate the
low-\textit{T} coherent Kondo state and the FL--NFL transition of
CeIrIn${_5}$. The most striking observation, however, is the
presence of a pseudogap-type precursor state preceding the SC in
this system, which is characterized by a change in the Hall
mobility. A microscopic comprehension of this precursor state
would be crucial; not only for understanding the electron pairing
in this system, but also in placing it in proper perspective with
respect to the high $T_c$ cuprates.

The authors are indebted to A. Gladun and C. Capan for useful
discussions. S.N. is supported by the Alexander von Humboldt
foundation. S.W. is partially supported by the EC through CoMePhS
517039. Z.F. acknowledges support through NSF-DMR-0710492. Work
at Los Alamos was performed under the auspices of the U.S.\
Department of Energy/Office of Science.

\end{document}